\documentclass[gmd, manuscript]{copernicus}

\usepackage[super]{nth}
\usepackage{hyperref}
\usepackage{soul}
\usepackage{lineno}

\begin{document}

\title{Towards physically consistent data-driven weather forecasting: Integrating data assimilation with equivariance-preserving deep spatial transformers}


\Author[1,2]{Ashesh}{Chattopadhyay}
\Author[2]{Mustafa}{Mustafa}
\Author[1,3]{Pedram}{Hassanzadeh}
\Author[4]{Eviatar}{Bach}
\Author[2]{Karthik}{Kashinath}

\affil[1]{Department of Mechanical Engineering, Rice University, Houston, TX, USA}
\affil[2]{Lawrence Berkeley National Laboratory, Berkeley, CA, USA}
\affil[3]{Department of Earth, Environmental and Planetary Sciences, Rice University, Houston, TX, USA}
\affil[4]{Department of Atmospheric and Oceanic Science and Institute for Physical Science and Technology, University of Maryland, College Park, USA}




\correspondence{Pedram Hassanzadeh (pedram@rice.edu)}

\runningtitle{Integrating data assimilation with equivariance-preserving deep spatial transformers}

\runningauthor{A Chattopadhyay et al.}

\received{}
\pubdiscuss{} 
\revised{}
\accepted{}
\published{}


\firstpage{1}

\maketitle
\nolinenumbers

\begin{abstract}
There is growing interest in data-driven weather prediction (DDWP), for example using convolutional neural networks such as U-NETs that are trained on data from models or reanalysis. Here, we propose 3 components to integrate with commonly used DDWP models in order to improve their physical consistency and forecast accuracy. These components are 1) a deep spatial transformer added to the latent space of the U-NETs to preserve a property called equivariance, which is related to correctly capturing rotations and scalings of features in spatio-temporal data, 2) a data-assimilation (DA) algorithm to ingest noisy observations and improve the initial conditions for next forecasts, and 3) a multi-time-step algorithm, which combines forecasts from DDWP models with different time steps through DA, improving the accuracy of forecasts at short intervals. To show the benefit/feasibility of each component, we use geopotential height at 500~hPa (Z500) from ERA5 reanalysis and examine the short-term forecast accuracy of specific setups of the DDWP framework. Results show that the equivariance-preserving networks (U-STNs) clearly outperform the U-NETs, for example improving the forecast skill by $45\%$. Using a sigma-point ensemble Kalman (SPEnKF) algorithm for DA and U-STN as the forward model, we show that stable, accurate DA cycles are achieved even with high observation noise. The DDWP+DA framework substantially benefits from large ($O(1000)$) ensembles that are inexpensively generated with the data-driven forward model in each DA cycle. The multi-time-step DDWP+DA framework also shows promises, e.g., it reduces the average error by factors of 2-3. These results show the benefits/feasibilities of these 3 components, which are flexible and can be used in a variety of DDWP setups. Furthermore, while here we focus on weather forecasting, the 3 components can be readily adopted for other parts of the Earth system, such as ocean and land, for which there is a rapid growth of data and need for forecast/assimilation.  
\end{abstract}

\introduction  
Motivated by improving weather and climate prediction, using machine learning (ML) for data-driven spatio-temporal forecasting of chaotic dynamical systems and turbulent flows has received substantial attention in recent years \citep[e.g.,][]{pathak2018model,vlachas2018data,dueben2018challenges,scher2018predicting,scher2019weather,chattopadhyay2020data,chattopadhyay2020deep,nadigareservoir,maulik2021reduced}. In fact, a few studies have already shown promising results with fully data-driven weather prediction (DDWP) models that are trained on variables representing the large-scale circulation obtained from numerical models or reanalysis products~\citep{scher2018toward,weyn2019can,weyn2020improving,chattopadhyay2019analog,chattopadhyay2018test,rasp2020weatherbench,arcomano2020machine,chantry2021opportunities,gronquist2021deep,watson2021machine,scherensemble}. These models leverage ML methods such as convolutional neural networks (CNNs) and/or recurrent neural networks (RNNs) that are trained on state variables representing the history of the spatio-temporal variability, and learn to predict the future states. 

The increasing interest \citep{schultz2021can,balaji2021climbing} in these DDWP models stems from the hope that they improve weather forecasting because of one or both of the following reasons: 1) trained on reanalysis data and/or data from high-resolution NWP models, these DDWP models may not suffer from some of the biases of physics-based, operational numerical weather prediction (NWP) models, and 2) the low computational cost of these DDWP models enables generating large ensembles for probabilistic forecasting \citep{weyn2020improving,weyn2021sub}. Regarding (1), while DDWP models trained on reanalysis data have skills for short-term predictions, so far they have not been able to outperform operational NWP models~\citep{weyn2020improving,arcomano2020machine,schultz2021can}. This might be, at least partly, due to the short training sets provided by around 40 years of high-quality reanalysis data \citep{rasp_2020_resnet}. There are a number of ways to tackle this problem, e.g., transfer learning could be used to blend data from low- and high-fidelity data/models \citep[e.g.,][]{ham2019deep,chattopadhyay2020super,rasp_2020_resnet}, and/or physical constraints could be incorporated into the often physics-agnostic ML models. \textit{The first contribution of this paper is to provide a framework for the latter, based on building physical properties called equivariances into convolutional architectures using deep spatial transformers.} The second contribution of this paper is to equip these DDWP models with data assimilation (DA), which is one of the key reasons behind the success of NWP models. Below, we further discuss the need for integrating DA and physical properties such as equivariances with DDWP models and briefly describe what has been already done in these areas in previous studies. 

Many of the DDWP models built so far are physics agnostic and learn the spatio-temporal evolution only from the training data, resulting sometimes in physically inconsistent predictions and inability to capture key invariants and symmetries of the underlying dynamical system, particularly when the training set is small~\citep{reichstein2019deep,chattopadhyay2019analog}. There are various approaches to incorporating some physical properties into the neural networks; for example, \citet{kashinath2021physics} have recently reviewed 10 approaches (with examples) for physics-informed ML in the context of weather/climate modeling. One popular approach, in general, is to enforce key conservation laws, symmetries, or some (or even all) of the governing equations through custom-designed loss functions~\citep[e.g.,][]{raissi2019physics,beucler2019achieving,daw2020physics,mohan2020embedding,thiagarajan2020designing,beucler2019enforcing}. 

Another approach--which has received less attention particularly in weather/climate modeling--is to enforce the appropriate symmetries, which are connected to conserved quantities through the Noether’s theorem~\citep{hanc2004symmetries}, inside the neural architecture. For instance, conventional CNN architectures enforce translational and rotational symmetries, which may not necessarily exist in the large-scale circulation; see \citet{chattopadhyay2019analog} for an example based on atmospheric blocking events and rotational symmetry. Indeed, recent research in the ML community has shown that preserving a more general property called ``equivariance'' can improve the performance of CNNs \citep{maron2018invariant,maron2019universality,cohen2019gauge}. Equivariance-preserving neural network architectures learn the existence of (or lack thereof) symmetries in the data rather than enforcing them \textit{a priori} and better track the relative spatial relationship of features~\citep{cohen2019gauge}. In fact, in their work on forecasting midlatitude extreme-causing weather patterns, \citet{chattopadhyay2019analog} have shown that capsule neural networks, which are equivariance-preserving \citep{sabour2017dynamic}, outperform conventional CNNs in terms of out-of-sample accuracy while requiring a smaller training set. Similarly, \citet{wang2020incorporating} have shown the advantages of equivariance-preserving CNN architectures in data-driven modeling of Rayleigh-B\'enard and ocean turbulence. More recently, using two-layer quasi-geostrophic turbulence as the test case, \citet{chattopadhyay2020deep} have shown that preserving equivariances related to translational, rotational, and scaling symmetry groups through a deep spatial transformer architecture \citep{jaderberg2015spatial} improves the accuracy and stability of the DDWP models without increasing the network's complexity or computational cost (which are drawbacks of capsule neural networks). Building on these studies, here our first goal is to develop a physically consistent, autoregressive DDWP model that preserves equivariance using a deep spatial transformer in an encoder-decoder U-NET architecture.   

DA is an essential component of modern weather forecasting \citep[e.g.,][]{kalnay2003atmospheric,carrassi2018data,lguensat2019data}. DA corrects the atmospheric state forecasted using a forward model (often a NWP model) by incorporating noisy and partial observations from the atmosphere (and other components of the Earth system), thus estimating a new corrected state of the atmosphere called ``analysis'', which serves as an improved initial condition for the forward model to forecast the future states. Most operational forecasting systems have their NWP model coupled to a DA algorithm that corrects the trajectory of the atmospheric states every $6$ h with observations from remote sensing and in-situ measurements. State-of-the-art DA algorithms use variational and/or ensemble-based approaches. The challenge with the former is computing the adjoint of the forward model, which involves high-dimensional, nonlinear partial differential equations \citep{penny2019strongly}. Ensemble-based approaches, which are usually variants of ensemble Kalman filter \citep[EnKF,][]{evensen1994sequential}, bypass the need for computing the adjoint but require generating a large ensemble of states that are each evolved in time using the forward model, which makes this approach computationally expensive \citep{hunt2007efficient,houtekamer2016review,kalnay2003atmospheric}. 

In recent years, there has been a growing number of studies at the intersection of ML and DA \citep{geer2021learning}. A few studies have aimed to use ML to accelerate/improve DA frameworks, for example by taking advantage of their natural connection~\citep{abarbanel2018machine,kovachki2019ensemble,grooms2021analog,hatfield2021building}. A few other studies have focused on using DA to provide suitable training data for ML from noisy/sparse observations \citep{brajard2020combining,brajard2020combiningSGS,tang2020deep,wikner2021using}. Others have integrated DA with a data-driven or hybrid forecast model for relatively simple dynamical systems \citep{hamilton2016ensemble,lguensat2017analog,lynch2019data,pawar2020data}. However, to the best of our knowledge, no study has yet integrated DA with a DDWP model. Here, our second goal is to present a DDWP+DA framework in which the DDWP is the forward model that efficiently provides a large, $O(1000)$, ensemble of forecasts for a sigma-point ensemble Kalman filter (SPEnKF) algorithm.

To provide proof-of-concepts for the DDWP model with equivariance-preserving encoder-decoder U-NET and the combined DDWP+DA framework, we use sub-daily $500$~hPa geopotential height (Z500) from the ECMWF Reanalysis 5 (ERA5) dataset  \citep{hersbach2020era5}. The DDWP model is trained on hourly, 6~h, or 12~h Z500 samples. The spatio-temporal evolution of Z500 is then forecasted from precise initial conditions using the DDWP model or from noisy initial conditions using the DDWP+SPEnKF framework. Our main contributions in this paper are three-fold, namely:
\begin{itemize}
    \item Introducing the equivariance-preserving encoder-decoder U-NET with a deep spatial transformer architecture for DDWP modeling and showing the advantages of this architecture over a conventional encoder-decoder U-NET. 
    
    \item Introducing the DDWP+DA framework, which leads to stable DA cycles without the need for any localization or inflation by taking advantage of the large forecast ensembles produced data drivenly using the DDWP model. 
    
    \item Introducing a novel multi-time-step framework for generating forecasts with short time steps that employs DA to ingest information from virtual observations produced using more accurate DDWP models that have longer time steps. This framework exploits the non-trivial dependence of the accuracy of autoregressive data-driven models on the time step size. 
\end{itemize}
The remainder of the paper is structured as follows. The data are described in Section~\ref{sec:data}. The encoder-decoder U-NET architecture with the deep spatial transformer and the SPEnKF algorithm are introduced in Section~\ref{sec:method}. Results are presented in Section~\ref{sec:results} and the Discussion and Summary are in Section~\ref{sec:discussion}.  

\section{Data}
\label{sec:data}
We use the ERA5 dataset from the WeatherBench repository \citep{rasp2020weatherbench}, where each global sample of Z500 at every hour is downsampled to a rectangular longitude-latitude $(x,y)$ grid of $32\times 64$. This coarse-resolution Z500 dataset from the WeatherBench repository has been used in a number of recent studies to perform data-driven weather forecasting~\citep{rasp2020weatherbench,rasp_2020_resnet}. Here, we use Z500 data from $1979$ to $2015$ for training, $2016$--$2017$ for validation, and $2018$ for testing.     

\section{Methods}\label{sec:method}
\subsection{The equivariance-preserving DDWP model: U-NET with a deep spatial transformer (U-STN)}
\label{sec:DL_forecast}

\begin{figure}[t]
\includegraphics[width = \linewidth]{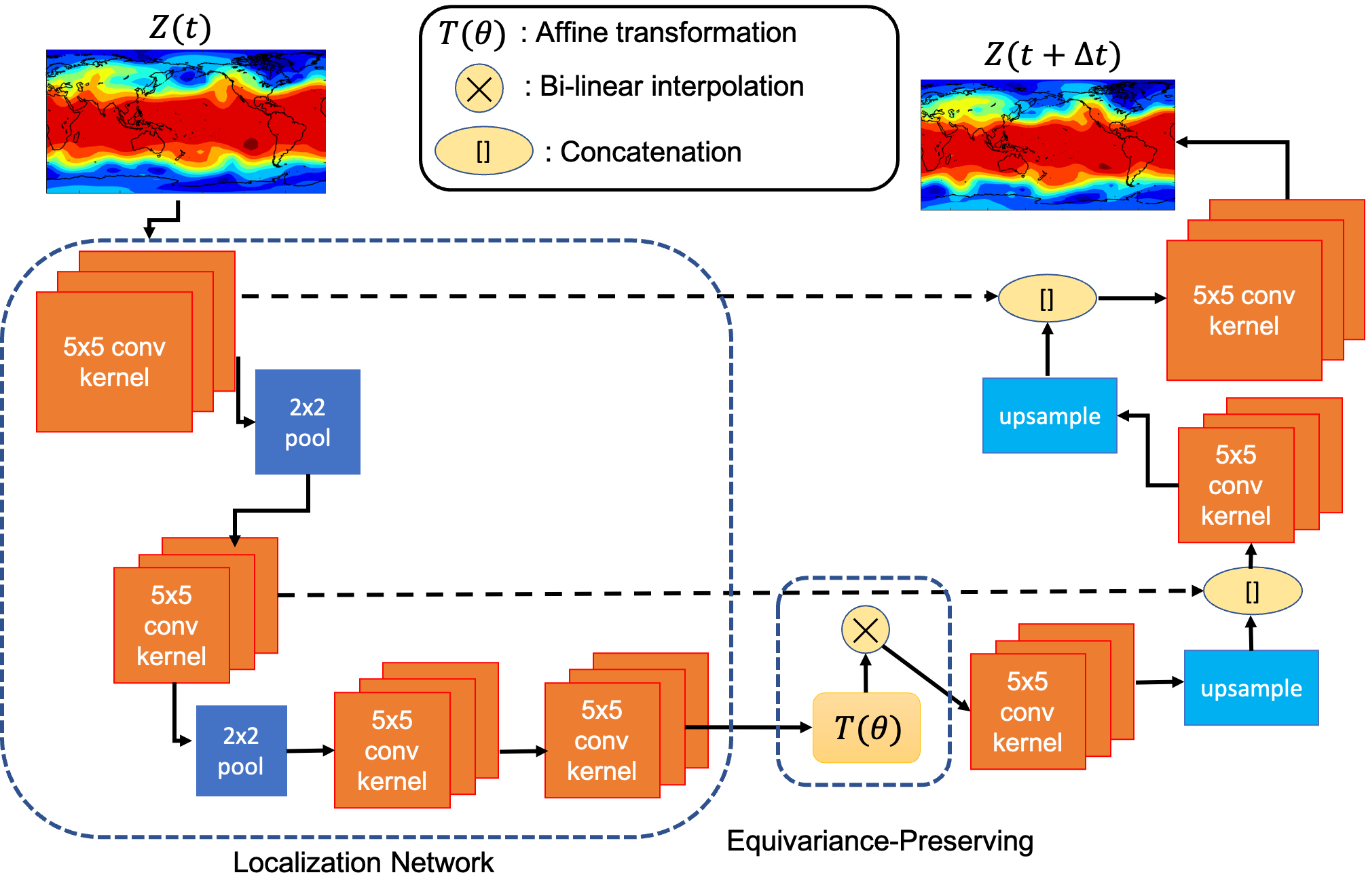}
\caption{\label{fig:U-STN} Architecture of U-STNx. The architecture is equivariance-preserving owing to the spatial transformer module implemented through the affine transformation, $T(\theta)$, along with the differentiable bi-linear interpolation kernel. The network integrates $Z(t)$ to $Z(t+\Delta t)$.}
\end{figure}

The DDWP models used in this paper are trained on Z500 data without access to any other atmospheric fields that might affect the atmosphere's spatio-temporal evolution. Once trained on past Z500 snapshots sampled at every $\Delta t$, the DDWP model takes Z500 at a particular time $t$ ($Z(t)$ hereafter) as the input and predicts $Z(t+\Delta t)$, which is then used as the input to predict $Z(t+2\Delta t)$, and this autoregressive process continues as needed. We use $\Delta t$ that is $1$, $6$, or $12$~h. The baseline DDWP model used here is a U-NET similar to the one used in \citet{weyn2020improving}. For the equivariance-preserving DDWP introduced here, the encoded latent space of the U-NET is coupled with a deep spatial transformer (U-STN hereafter). The preservation of equivariance enables the U-STN to track rotation and stretching of the synoptic- and larger-scale patterns, and is expected to improve the forecast of the spatio-temporal evolution of the midlatitude Rossby waves and their nonlinear breakings. In this section, we briefly discuss the U-STN architecture, which is schematically shown in Fig.~\ref{fig:U-STN}. Note that from now on, ``x'' in U-STNx (and U-NETx) indicates the $\Delta t$ that is used, e.g., U-STN6 uses $\Delta t=6$~h.     

\subsubsection{Localization network or encoding block of U-STN}
\label{encode_block}
The network takes in an input snapshot of Z500, $Z(t)^{32 \times 64}$, as initial condition and projects it onto a low-dimensional encoding space via a U-NET convolutional encoding block. This encoding block performs three convolutions (without changing the spatial dimensions), the first two of which are followed by max-pooling. The convolutions inside the encoder block account for Earth's longitudinal periodicity by performing circular convolutions \citep{schubert2019circular} on each feature map inside the encoder block. The encoded feature map, which is the output of the encoding block and consists of the reduced $Z$ and co-ordinate system, $\tilde{Z}^{8 \times 16}$ and $(x^o_i,y^o_i)$ where $i=1,2 \dots 8 \times 16$, is sent to the spatial transformer module described below. 

\subsubsection{Spatial transformer module}
\label{STN}
The spatial transformer applies an affine transformation $T(\theta)$ to the reduced co-ordinate system $(x^o_i,y^o_i)$ to obtain a new transformed co-ordinate system $(x^s_i,y^s_i)$:
\begin{align}
\label{eq:affine}
    \begin{bmatrix}
    x_i^s\\
    y_i^s\\
    \end{bmatrix}
    =T(\theta) 
    \begin{bmatrix}
     x_i^o\\
    y_i^o\\
    1 
    \end{bmatrix},
\end{align}
where
\begin{align}
  T\left(\theta\right)=
        \begin{bmatrix}
       \theta_{11} & \theta_{12} & \theta_{13}\\
        \theta_{21} & \theta_{22} & \theta_{23} 
\end{bmatrix}.
\end{align}
The parameters $\theta$ are learnt through backpropagation. A differentiable sampling kernel (a bi-linear interpolation kernel in this case) is then used to transform $\tilde{Z}^{8 \times 16}$, which is on the old co-ordinate system $(x^o_i,y^o_i)$, into $\bar{Z}^{8 \times 16}$, which is on the new co-ordinate system $(x^s_i,y^s_i)$. The spatial transformer module ensures that the latent space that is encoded is equivariance-preserving~\citep{esteves2018learning}, which enables the U-STN to correctly tracks relative positions (rotation and scaling) of the multi-scale features inherently present in turbulent flows~\citep{wang2020incorporating,chattopadhyay2020deep}. 

\subsubsection{Decoding block}
The decoding block is a series of deconvolution layers (convolution with zero-padded upsampling) concatenated with the corresponding convolution outputs from the encoder part of the U-NET. The decoding blocks bring the encoded equivariance-preserving latent space $\bar{Z}^{8 \times 16}$ back into the original dimension and co-ordinate system at time $t+ \Delta t$, thus outputting $Z(t+\Delta t)^{32 \times 64}$. The concatenation of the encoder and decoder convolution outputs allows the architecture to better learn the features in the small-scale dynamics of Z500 \citep{weyn2020improving}. 

The loss function $L$ to be minimized is 
\begin{eqnarray}
L(\lambda)=\frac{1}{(N+1)}\sum_{t=0}^{t=N\Delta t}||\left(Z(t+\Delta t)-\text{U-STNx}\left(Z(t),\lambda\right)\right)||_2^2,
\end{eqnarray}
where $N$ is the number of training samples, $t=0$ is the start time of the training set, and $\lambda$ represents the parameters of the network that are to be trained (in this case, the weights, biases, and $\theta$ of U-STNx). In both encoding and decoding blocks, the ReLU activation functions are used. The number of convolutional kernels ($32$ in each layer), size of each kernel ($5 \times 5$), and the learning rate ($\alpha=3 \times 10^{-4}$) have been chosen after extensive search. All codes for these networks (as well as DA) have been made publicly available on GitHub (see the Code Availability statement).

Note that without the transformer module, $\bar{Z}=\tilde{Z}$, and the network becomes a U-NET. Also, we highlight that here, we are focusing on preserving the SO(3) equivariance group that includes translation, rotation, and scaling, because those are the ones that matter the most for the synoptic patterns on a 2D plane. Other transformations and equivariance groups could be similarly included \citep{wang2020incorporating}. Furthermore, here we focus on an architecture with a transformer that acts only on the latent space. More complex architectures, with transformations like Eq.~(\ref{eq:affine}) after every convolution layer can be used too \citep{de2020gauge,wang2020incorporating}. Our preliminary exploration shows that for this work, the one spatial transformer module  applied on the latent space of the U-NET yields sufficiently superior performance (over the baseline, U-NET), but further exhaustive explorations should be conducted in future studies to find the best performing architecture for each application. 



\subsection{Data assimilation algorithm and coupling with DDWP}
\label{sec:DA}
For DA, we employ the SPEnKF algorithm, which unlike the EnKF algorithm, does not use random perturbations to generate an ensemble but rather uses an unscented transformation to deterministically find an optimal set of points called sigma points \citep{ambadan2009sigma}. The SPEnKF algorithm has been shown to outperform EnKF on particular test cases for both chaotic dynamical systems and ocean dynamics \citep{tang2014practical} although whether it is always superior to EnKF is a matter of active research \citep{hamill2009comments} and beyond the scope of this paper. Our DDWP+DA framework can use any ensemble-based algorithm.  

\begin{figure}[t]
\includegraphics[width = \linewidth]{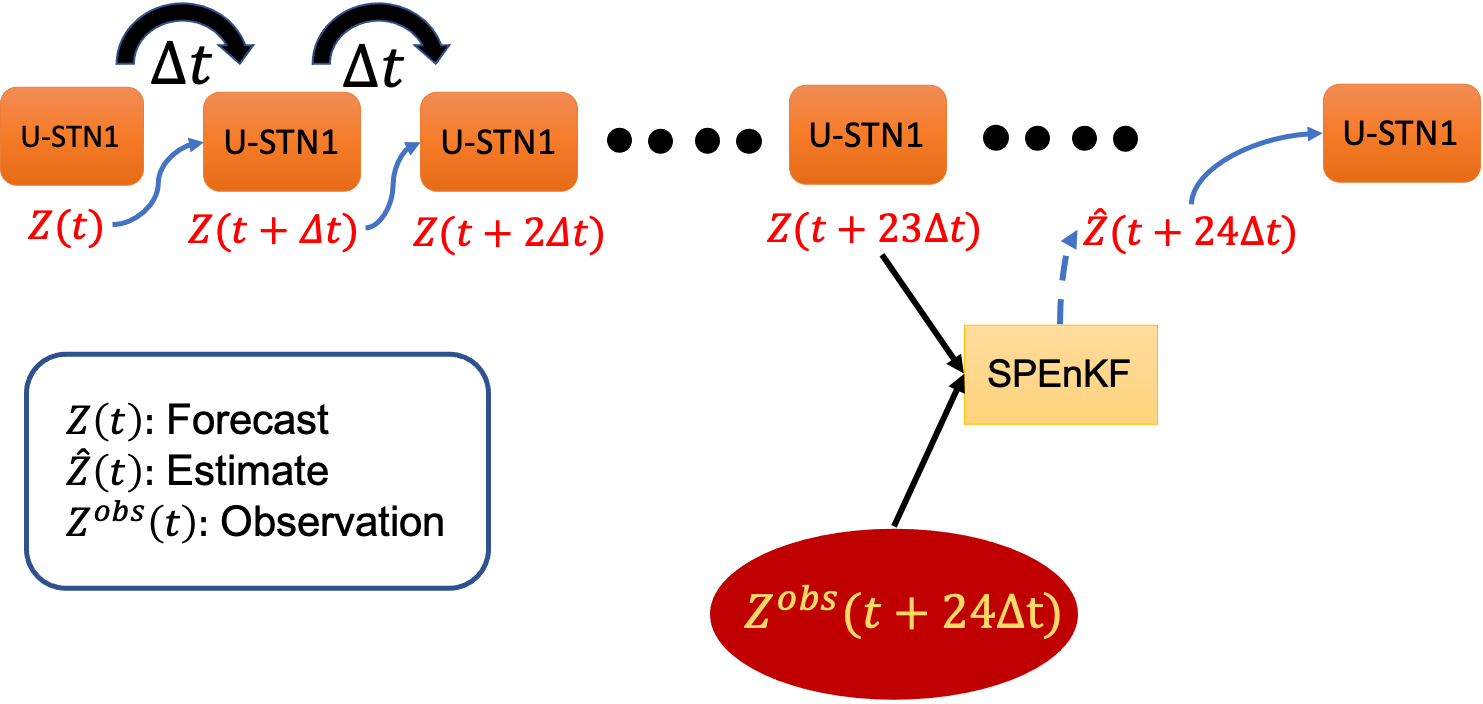}
\caption{\label{fig:DL_DA_cartoon} The framework for a synergistic integration of a DA algorithm (SPEnKF) with a DDWP (U-STN1). Once the DDWP+DA framework is provided with a noisy $Z(t)$, it uses U-STN1 to autoregressively predict $Z(t+23\Delta t)$. A large ensemble is then generated using Eq.~\eqref{eq:ensemble_gen}, and for each member $k$, $Z^k_\text{ens}(t+24\Delta t)$ is predicted using U-STN1. Following that, an SPEnKF algorithm assimilates a noisy observation at the $\nth{24}$~h to provide the estimate (analysis) state of Z500, $\hat{Z}(t+24 \Delta t)$. U-STN1 then uses this analysis state as the new initial condition and evolves the state in time, with DA occurring every $24$ hours. }
\end{figure}

In the DDWP+DA framework, shown schematically in Fig.~\ref{fig:DL_DA_cartoon}, the forward model is a DDWP, which is chosen to be U-STN1 and denoted as $\mathbf{\Psi}$ below. We use $\sigma_\text{obs}$ for the standard deviation of the observation noise, which in this paper is either $\sigma_\text{obs}=0.5 \sigma_{Z}$ or $\sigma_\text{obs}=\sigma_{Z}$, where $\sigma_{Z}$ is the standard deviation of $Z500$ over all grid points and over all years between $1979$--$2015$. Here, we assume that the noisy observations are assimilated every $24$~h (again, the framework can be used with any DA frequency, such as $6$~h, which is used commonly in operational forecasting).  

We start with a noisy initial condition $Z(t)$, and use U-STN1 to autoregressively (with $\Delta t=1$~h) predict the next time steps, $Z(t+\Delta t)$, $Z(t+2\Delta t)$, $Z(t+3\Delta t)$, up to $Z(t+23\Delta t)$. For a $D$-dimensional system (i.e., $Z \in \cal{R}^{D}$), the optimal number of ensemble members for SPEnKF is $2D + 1$ \citep{ambadan2009sigma}. Because here $D=32\times64$, $4097$ ensemble members are needed. While this is a very large ensemble size if the forward models is a NWP (operationally, $\sim 50-100$ members are used), the DDWP can inexpensively generate $O(1000)$ ensemble members, a major advantage of DDWP as a forward model that we will discuss later in Section~\ref{sec:discussion}. 

To do SPEnKF, an ensemble of states at the $2\nth{3}$ hour of each DA cycle ($24$~h is one DA cycle) is generated using a symmetric set of sigma points \citep{julier2004unscented} as
\begin{eqnarray}
\begin{aligned}
\label{eq:ensemble_gen}
Z^{i}_\text{ens}(t+23 \Delta t)&=Z(t+23 \Delta t)- A_i, \\
Z^{j}_\text{ens}(t+23 \Delta t)&=Z(t+23 \Delta t)+ A_j, \\
Z^{0}_\text{ens}(t+23 \Delta t)&=Z(t+23 \Delta t),
\end{aligned}
\end{eqnarray}
where $i,j \in \left[1,2,\cdots D=32\times64\right]$ and $0$ are indices of the $2D+1$ ensemble members. Vectors $A_i$ and $A_j$ are columns of matrix $\mathbf{A}=\mathbf{U \sqrt S  U^T}$, where $\mathbf{U}$ and $\mathbf{S}$ are obtained from the singular value decomposition of the analysis covariance matrix $\mathbf{P_a}$, i.e., $\mathbf{P_a=USV}^T$. The $D\times D$ matrix $\mathbf{P_a}$ is either available from the previous DA cycle (see Eq.~\eqref{eq:final_P} below) or is initialized as an identity matrix at the beginning of DA. Note that here, we generate the ensemble at one $\Delta t$ before the next DA; however, the ensembles can be generated at any time within the DA cycle and carried forward although that would increase the computational cost of the framework. We have explored generating the ensembles at $t+0\Delta t$ (i.e., the beginning) but did not find any improvement over Eq.~\eqref{eq:ensemble_gen}.

Once the ensembles are generated via Eq.~\eqref{eq:ensemble_gen}, every ensemble member is fed into $\mathbf{\Psi}$ to predict an ensemble of forecasted states at $t+ 24 \Delta t$:
\begin{eqnarray}
\label{eq:forward_model}
Z^k_\text{ens}(t+24\Delta t) = \mathbf{\Psi}\left(Z^k_\text{ens}(t+23\Delta t)\right), 
\end{eqnarray}
where $k \in \left[-D, -D+1, \cdots , D-1, D\right]$. In general, the modeled observation is $\mathbf{H}\left(\left<Z_\text{ens}^{k}(t+24 \Delta t)\right>,\epsilon(t)\right)$, where $\mathbf{H}$ is the observation operator and $\epsilon(t)$ is the Gaussian random process with standard deviation $\sigma_\text{obs}$ that represents the observation noise. $\left< . \right>$ denotes ensemble averaging. In this paper, we assume that $\mathbf{H}$ is the identity matrix while we acknowledge that in general, it could be a nonlinear function. The SPEnKF algorithm can account for such complexity, but here, to provide a proof-of-concept, we have assumed that we can observe the state, although with a certain level of uncertainty. With $\mathbf{H}=\mathbf{I}$, the background error covariance matrix $\mathbf{P_b}$ becomes
\begin{eqnarray}
\label{eq:P_b}
\mathbf{P_b} = \mathbf{E}\left[ \left(Z^k_\text{ens}(t+24\Delta t)-\left<Z_\text{ens}^{k}(t+24 \Delta t)\right>\right) \left(Z^k_\text{ens}(t+24\Delta t)-\left<Z_\text{ens}^{k}(t+24 \Delta t)\right>\right)^T \right]/(4D+1), 
\end{eqnarray}
where $\left[.\right]^T$ denotes the transpose operator and $\mathbf{E}[.]$ denotes the expectation operator. The innovation covariance matrix is defined as:
\begin{eqnarray}
\label{eq:innovation}
\mathbf{C}=\mathbf{P_b}+\mathbf{R},
\end{eqnarray}
where the observation noise matrix $\mathbf{R}$ is a constant diagonal matrix of the variance of observation noise, i.e., $\sigma^2_\text{obs}$. Finally, we compute the cross-covariance matrix:
\begin{eqnarray}
\label{eq:P_ab}
\mathbf{P_{ab}} = \mathbf{E}\left[ \left(Z^k_\text{ens}(t+24\Delta t)-\left<Z^k_\text{ens}(t+24\Delta t)\right> \right) \left(Z^k_\text{ens}(t+24\Delta t)-\left<Z_\text{ens}^{k}(t+24 \Delta t)\right>\right)^T \right]/(4D+1).
\end{eqnarray}
The Kalman gain matrix is then given by
\begin{eqnarray}
\label{eq:K}
\mathbf{K=P_{ab}C}^{-1},
\end{eqnarray}
and the estimated (analysis) state $\hat{Z}(t+24\Delta t)$ is calculated as
\begin{eqnarray}
\label{eq:estimated_Z}
\hat{Z}(t+24\Delta t)=\left<Z(t+24\Delta t)\right>- \mathbf{K} \left(\left<Z_\text{ens}^{k}(t+24 \Delta t)\right>-Z^\text{obs}(t+24\Delta t)\right),
\end{eqnarray}
where $Z^\text{obs}(t+24\Delta t)$ is the noisy observed Z500 at $t+24 \Delta t$; i.e., ERA5 value at each grid point plus random noise drawn from $\mathcal{N}(0,\sigma_\text{obs})$. The analysis error covariance matrix is updated as
\begin{eqnarray}
\label{eq:final_P}
\mathbf{P_a = KP_bK}^T.
\end{eqnarray}
The estimated state $\hat{Z}(t+24\Delta t)$ becomes the new initial condition to be used by U-STN1 and the updated $\mathbf{P_a}$ is used to generate the ensembles in Eq.~(\ref{eq:ensemble_gen}) after another $23$~h for the next DA cycle. 

Finally, we remark that often with low ensemble sizes, the background covariance matrix, $\mathbf{P_b}$ (Eq.~(\ref{eq:P_b})), suffers from spurious correlations which are corrected using localization and inflation strategies~\citep{hunt2007efficient,asch2016data}. However, due to the large ensemble size used here (with $4097$ ensemble members that are affordable because of the computationally inexpensive DDWP forward model) we do not need to perform any localization or inflation on $\mathbf{P_b}$ to get stable DA cycles as shown in the next section.

\section{Results}
\label{sec:results}
\subsection{Performance of physically consistent DDWP: Noise-free initial conditions (no DA)}
\label{sec: results_withoutDA}

\begin{figure}[t]
\includegraphics[width = 0.75\linewidth]{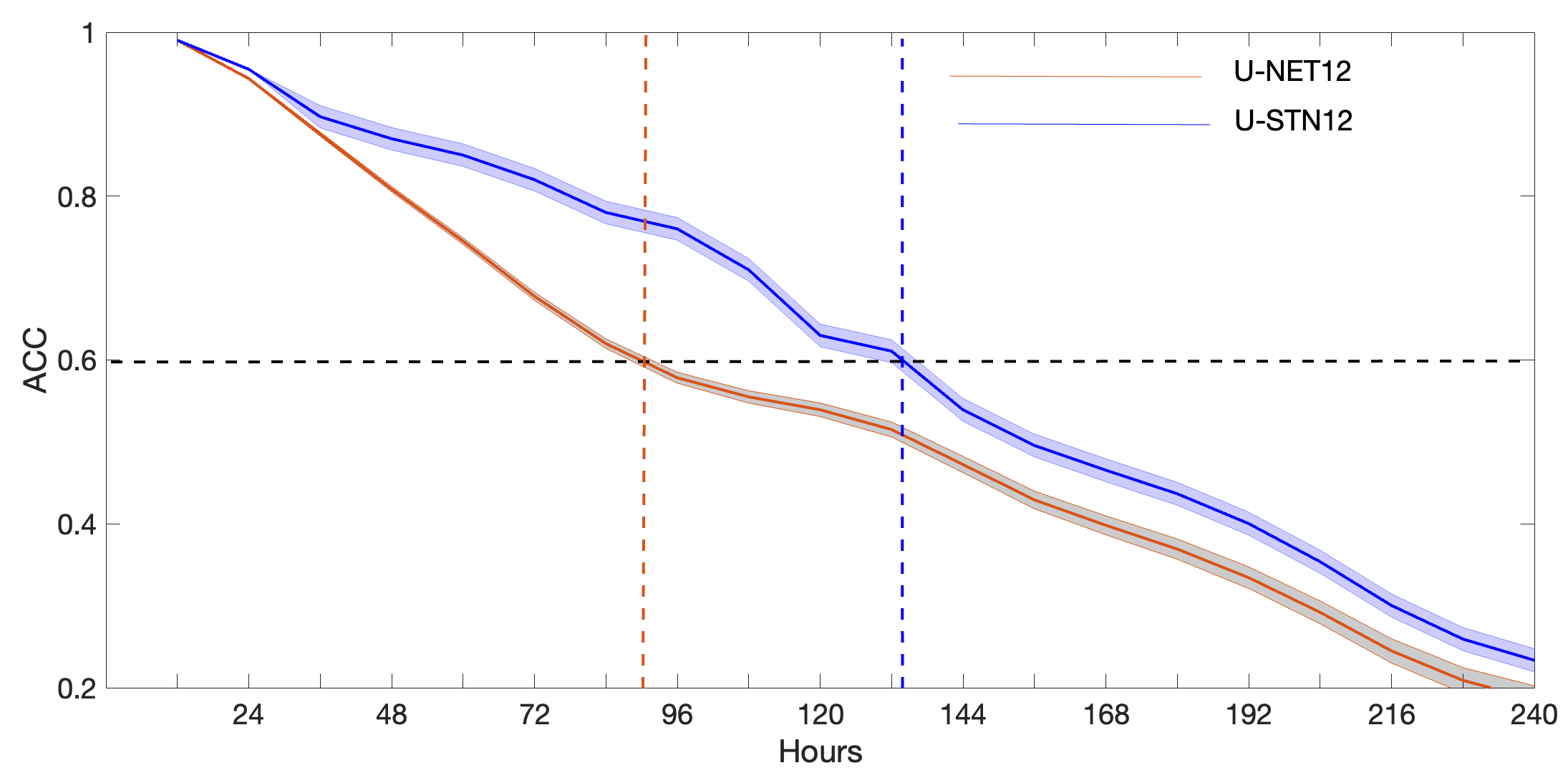}
\caption{\label{fig:12hrly_ACC} Anomaly correlation coefficient (ACC) calculated between Z500 anomalies of ERA5 and Z500 anomalies predicted using U-STN12 or U-NET12 from 30 noise-free, random initial conditions. The solid lines and the shadings show the mean and the standard deviation over the 30 initial conditions.}
\end{figure}

First, we show the gain from preservation of the equivariances by comparing the performance of a U-STN and a conventional U-NET, whose only difference is in the use of the spatial transformer module in the former. Using U-STN12 and U-NET12 as representatives of these architectures, Fig.~\ref{fig:12hrly_ACC} shows the anomaly correlation coefficients (ACCs) between the predictions from U-STN12 or U-NET12 and the truth (ERA5) for 30 noise-free, random initial conditions. ACC is computed every 12~h as the correlation coefficient between the predicted Z500 anomaly and the Z500 anomaly of ERA5, where anomalies are derived by removing the $1979$-$2015$ time mean of Z500 of the ERA5 dataset. U-STN12 clearly outperforms U-NET12, most notably after 36~h, reaching ACC=0.6 after around 132~h, a $45\%$ (1.75~day) improvement over U-NET12, which reaches ACC=0.6 after around 90~h.    

\begin{figure}[t]
\includegraphics[width = \linewidth]{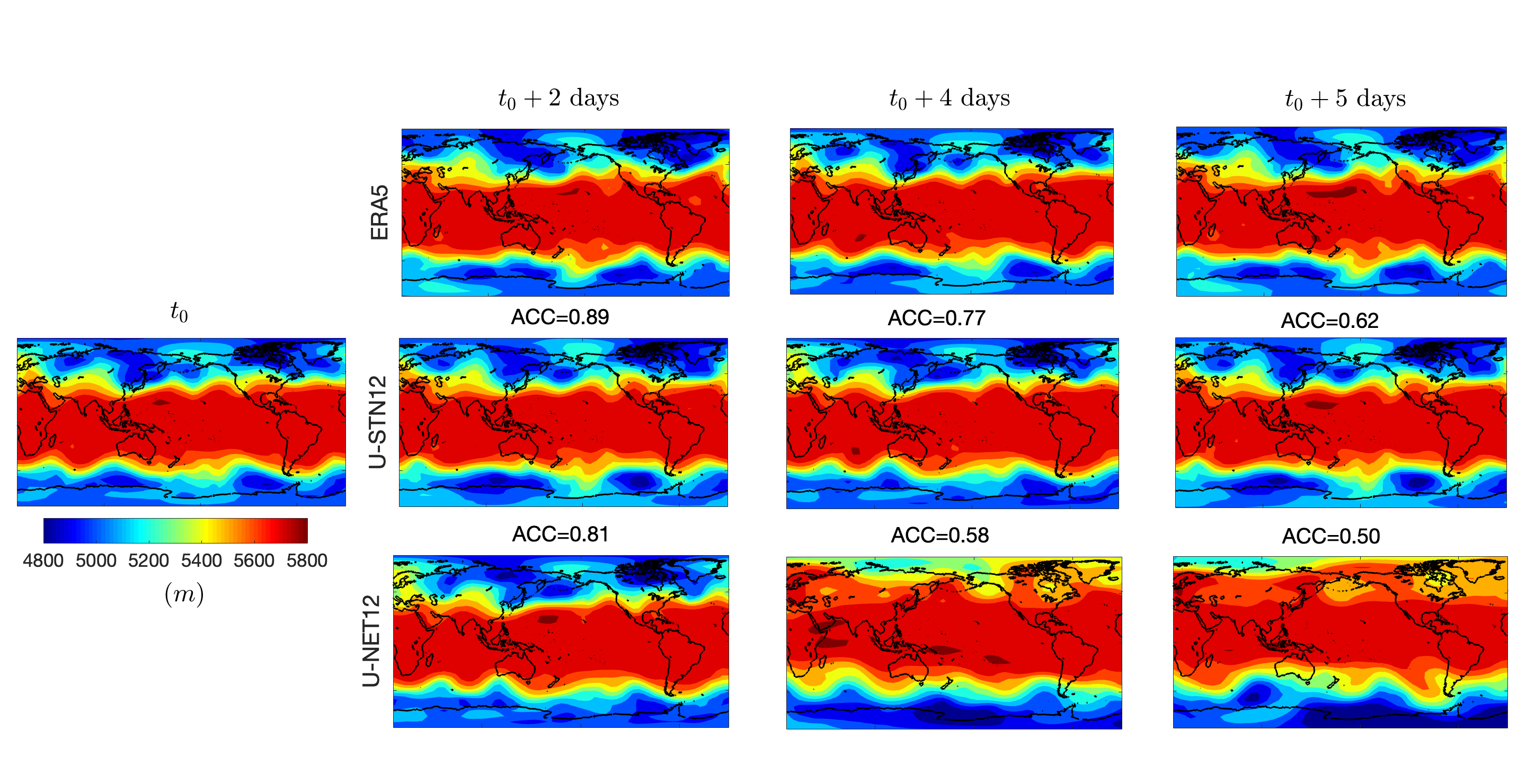}
\caption{\label{fig:12hrly_Unet_DD} Examples of the spatio-temporal evolution of Z500 predicted from a noise-free initial condition ($t_0$) using U-STN12 and U-NET12, and compared with the truth from ERA5. For the predicted patterns, the anomaly correlation coefficient (ACC) is shown above each panel (see the text for details). }
\end{figure}

To further see the source of this improvement, Fig.~\ref{fig:12hrly_Unet_DD} shows the spatio-temporal evolution of Z500 patterns from an example of prediction using U-STN12 and U-NET12. Comparing with the truth (ERA5), U-STN12 can better capture the evolution of the large-amplitude Rossby waves and the weavebreaking events compared to U-NET12; e.g., see the patterns over Central Asia, Southern Pacific Ocean, and Northern Atlantic Ocean on days 2-5. As discussed before, improvements in capturing of wavebreaking events, which involve rotation of synoptic features, are expected from an equivariance-preserving network such as U-STN. Furthermore, on days 4 and 5, the predictions from U-NET12 have substantially low Z500 values in the high latitudes of the Southern Hemisphere, showing signs of unphysical drifts.       

Overall, the results of Figs.~\ref{fig:12hrly_ACC} and~\ref{fig:12hrly_Unet_DD} show the advantages of using equivariance-preserving U-STNs in DDWP models. Note that while here we show results with $\Delta t =12$~h, similar improvements are seen with $\Delta t =1$~h and $\Delta t = 6$~h (see section~\ref{sec:VirtualObs}). Furthermore, to provide a proof-of-concept for the U-STN, in this paper we focus on Z500 (representing the large-scale circulation) as the only state variable to be learnt and predicted. Even without access to any other information (for example about small scales), the DDWP model can provide skillful forecasts for some time, consistent with earlier findings with the multi-scale Lorenz 96 system \citep{dueben2018challenges,chattopadhyay2020data}. More state variables can be easily added to the framework, which is expected to extend the forecast skills, based on previous work with U-NETs \citep{weyn2020improving}.

\subsection{Performance of the DDWP+DA framework: noisy initial conditions and assimilated observations}
\label{sec:DD_with1DA}
To provide a proof-of-concept for the DDWP+DA framework, we use U-STN1 as the DDWP model and SPEnKF as the DA algorithm, as described in Section~\ref{sec:DA}. In this U-STN1+SPEnKF setup, the initial conditions for predictions are noisy observations and every $24$~h, noisy observations are assimilated to correct the forecast trajectory (as mentioned before, noisy observations are generated by adding random noise from $\mathcal{N}(0,\sigma_{obs})$ to the Z500 of ERA5). 

\begin{figure}[t]
\includegraphics[width = \linewidth]{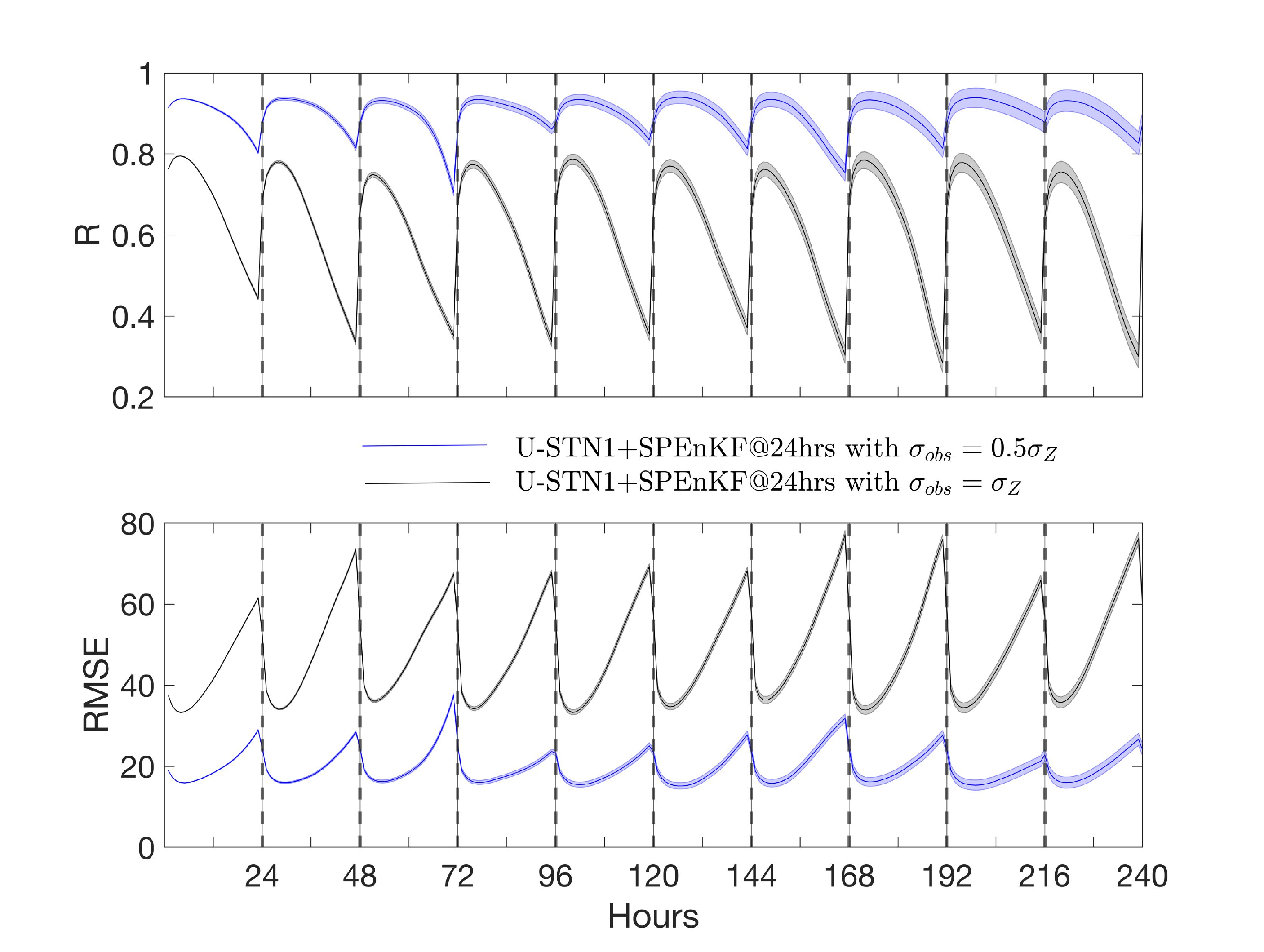}
\caption{\label{fig:DL_1DA} The top (bottom) panel shows R (RMSE, in meters) between noise-free data from ERA5 and the forecasts from U-STN1+SPEnKF for two levels of observation noise. Predictions are started from 30 random noisy observations. The lines (shading) show the mean (standard deviation) of the 30 forecasts. Noisy observations are assimilated every $24$~h (indicated by black, dashed vertical lines).}
\end{figure}

In Fig.~\ref{fig:DL_1DA}, for 30 random initial conditions and two noise levels ($\sigma_{obs}=0.5$ or $1\sigma_Z$), we report the spatially averaged root-mean-squared-error (RMSE) and the correlation coefficient (R) of the forecasted full Z500 fields as compared to the truth, i.e., the (noise-free) Z500 fields of ERA5. For both noise levels, we see that within each DA cycle, the forecast accuracy decreases between $0$ and $23$~h until DA with SPEnKF occurs at the $\nth{24}$ hour wherein information from the noisy observation is assimilated to improve the estimate of the forecast at the $\nth{24}$ hour. This estimate acts as the new improved initial condition to be used by U-STN1 to data drivenly forecast future time steps. In either case, the RMSE and R remain below 30~m (80~m) and above 0.7 (0.3) with $\sigma_{obs}=0.5\sigma_Z$ ($\sigma_{obs}=1\sigma_Z$) for the first 10 days. The main point here is not the accuracy of the forecast (which as mentioned before, could be further extended, for example by adding more state variables), but the stability of the U-STN1+SPEnKF framework (without localization/inflation), which even with the high noise level, can correct the trajectory, and increase R from $\sim 0.3$ to $0.8$ in each cycle. Although not shown in this paper, the U-STN1+SPEnKF framework remains stable beyond the 10 days and shows equally good performance for longer periods of time. 

One last point to make here is that within each DA cycle, the maximum forecast accuracy is not at when DA occurs, but 3-4~h later (this is most clearly seen for the case with $\sigma_{obs}=1\sigma_Z$ in Fig.~\ref{fig:DL_1DA}). The reason behind the further improvement of the performance after DA is the de-noising capability of neural networks~\citep{xie2012image}. Since the U-STN1 model has been trained on noise-free ERA5 data, it possesses inherent de-noising properties, which enable the model to further improve the forecast for a few hours after the noisy observation is assimilated.


\subsection{DDWP+DA with virtual observations: A multi-time-step framework}
\label{sec:VirtualObs}

\begin{figure}[t]
\includegraphics[width = \linewidth]{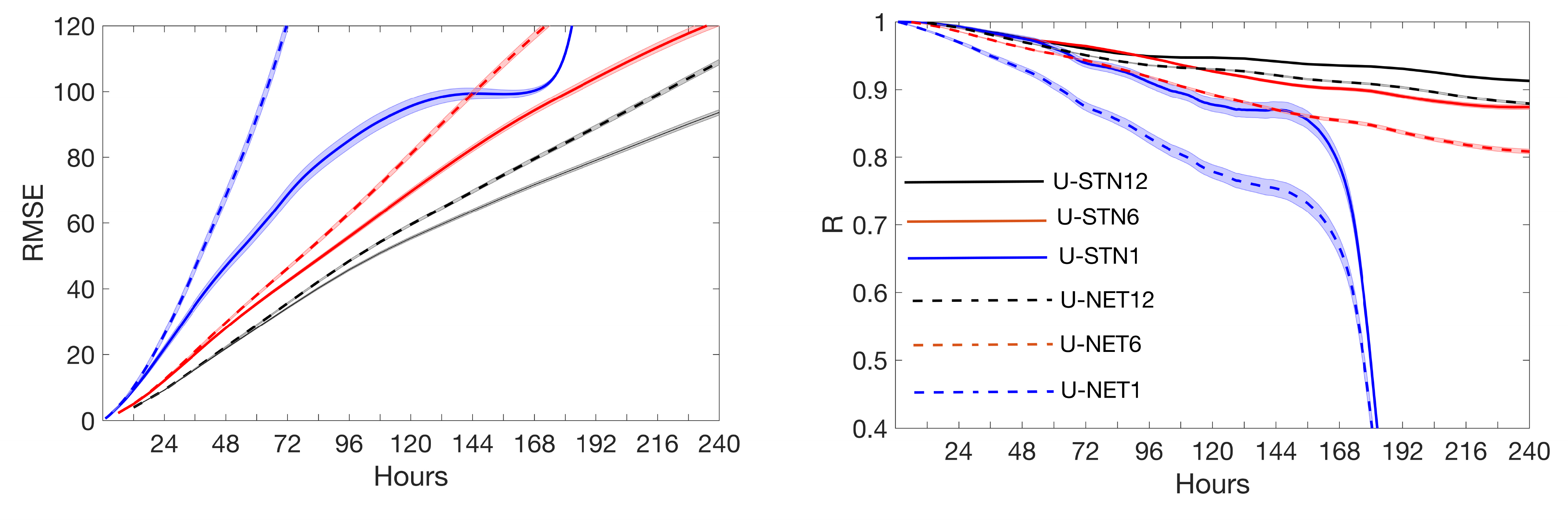}
\caption{\label{fig:all_correlation} The left (right) panel shows RMSE (R) between noise-free data from ERA5 and the forecasts from U-STNx or U-NETx from 30 random, noise-free initial conditions. No DA is used here. RMSE is in meters. The lines (shading) show the mean (standard deviation) of the 30 forecasts.}
\end{figure}

One might wonder how the performance of the DDWP model (with or without DA) depends on $\Delta t$. Figure~\ref{fig:all_correlation} compares the performance of U-STNx as well as U-NETx for $\Delta t=1$, $6$, and $12$~h for 30 random noise-free initial conditions (no DA). It is clear that the DDWP models with larger $\Delta t$ outperform the ones with smaller $\Delta t$; i.e., in terms of forecast accuracy, U-STN12 $>$ U-STN6 $>$ U-STN1. This trends holds true for both U-STNx and U-NETx, while as discussed before, for the same $\Delta t$, the equivariance-preserving U-STN outperforms the U-NET.  

This dependence on $\Delta t$ might seem counter-intuitive as it is opposite of what one sees in numerical models, whose forecast errors decrease with smaller time steps. The increase in the forecast errors of these DDWP models when $\Delta t$ is decreased is likely due to the non-additive nature of the error accumulation of these autoregressive models. The data-driven models have some degree of generalization error (for out-of-sample prediction), and every time the model is invoked to predict the next time step, this error is accumulated. For neural networks, this accumulation is not additive and propagates nonlinearly during the autoregressive prediction. Currently, these error propagations are not understood well enough to build a rigorous framework for estimating the optimal $\Delta t$ for data-driven, autoregressive forecasting; however, this behavior has been reported in other studies on nonlinear dynamical systems and can be exploited to formulate multi-time-step data-driven models; see~\citep{liu2020hierarchical} for an example (though without DA). 

\begin{figure}[t]
\includegraphics[width = \linewidth]{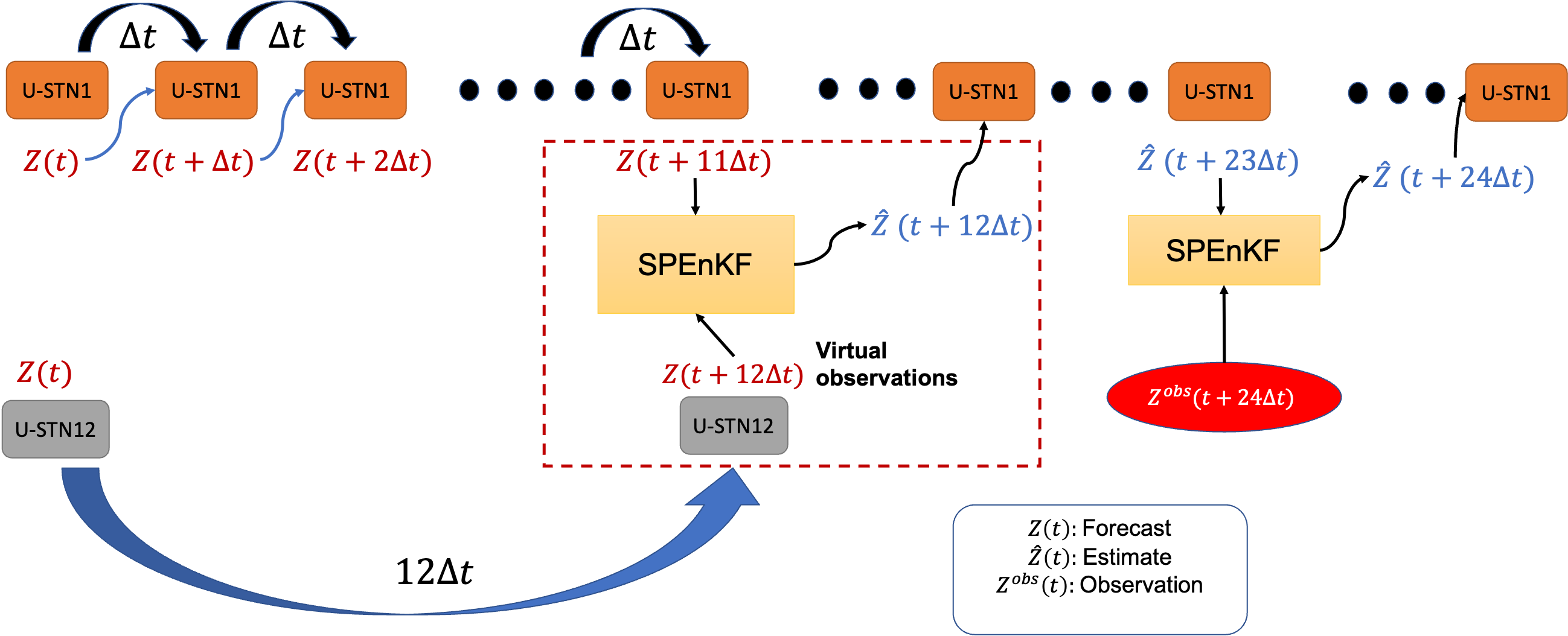}
\caption{\label{fig:virtual_obs_frame} Schematic of the multi-time-step DDWP+DA framework. The U-STN12 model provides forecasts every 12~h, which are assimilated as virtual observations using SPEnKF into the U-STN1+SPEnKF framework that has a 24~h DA cycle for assimilating noisy observations. At $\nth{12}$ hours, the U-STN12 forecasts are more accurate than those from the U-STN1 model, enabling the framework to improve the prediction accuracy every $\nth{12}$ hour, thereby improving the initial condition used for the next forecasts before DA with noisy observations (every $24$~h).}
\end{figure}

Based on the trends seen in Fig.~\ref{fig:all_correlation}, we propose a novel idea for a multi-time-step DDWP+DA framework, in which the forecasts from the more accurate DDWP with larger $\Delta t$ are incorporated as virtual observations, using DA, into the forecasts of the less accurate DDWP with smaller $\Delta t$, thus providing overall more accurate short-term forecasts. Figure~\ref{fig:virtual_obs_frame} shows a schematic of this framework for the case where the U-STN12 model provides the virtual observations that are assimilated using the SPEnKF algorithm in the middle of the 24~h DA cycles into the hourly forecasts from U-STN1. At $\nth{24}$~hours, noisy observations are assimilated using the SPEnKF algorithm as before. 

\begin{figure}[t]
\includegraphics[width = \linewidth]{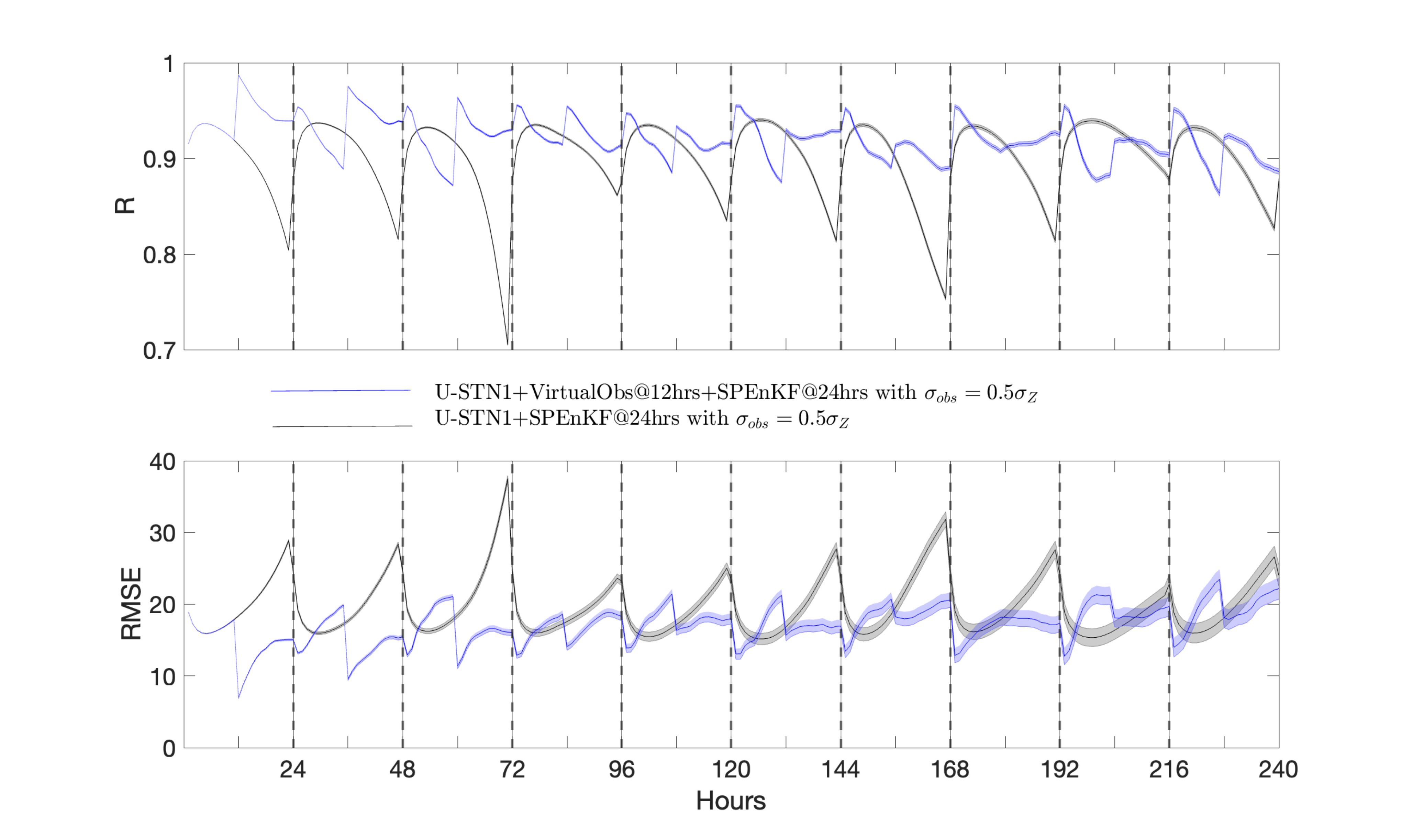}
\caption{\label{fig:DL_2DA_0.5} Performance of the the multi-time-step U-STNx+SPEnKF framework (with virtual observations at the $\nth{12}$~hour of every $24$ h DA cycle) compared to that of the U-STN+SPEnKF framework for the case with $\sigma_\text{obs}=0.5\sigma_{Z}$. The top (bottom) panel show R (RMSE in meters). The black, dashed vertical lines indicate DA of noisy observations at every $24$~h. Forecasts are started from 30 random, noisy initial conditions. The lines (shading) show the mean (standard deviation) of the 30 forecasts.}
\end{figure}

\begin{figure}[t]
\includegraphics[width = \linewidth]{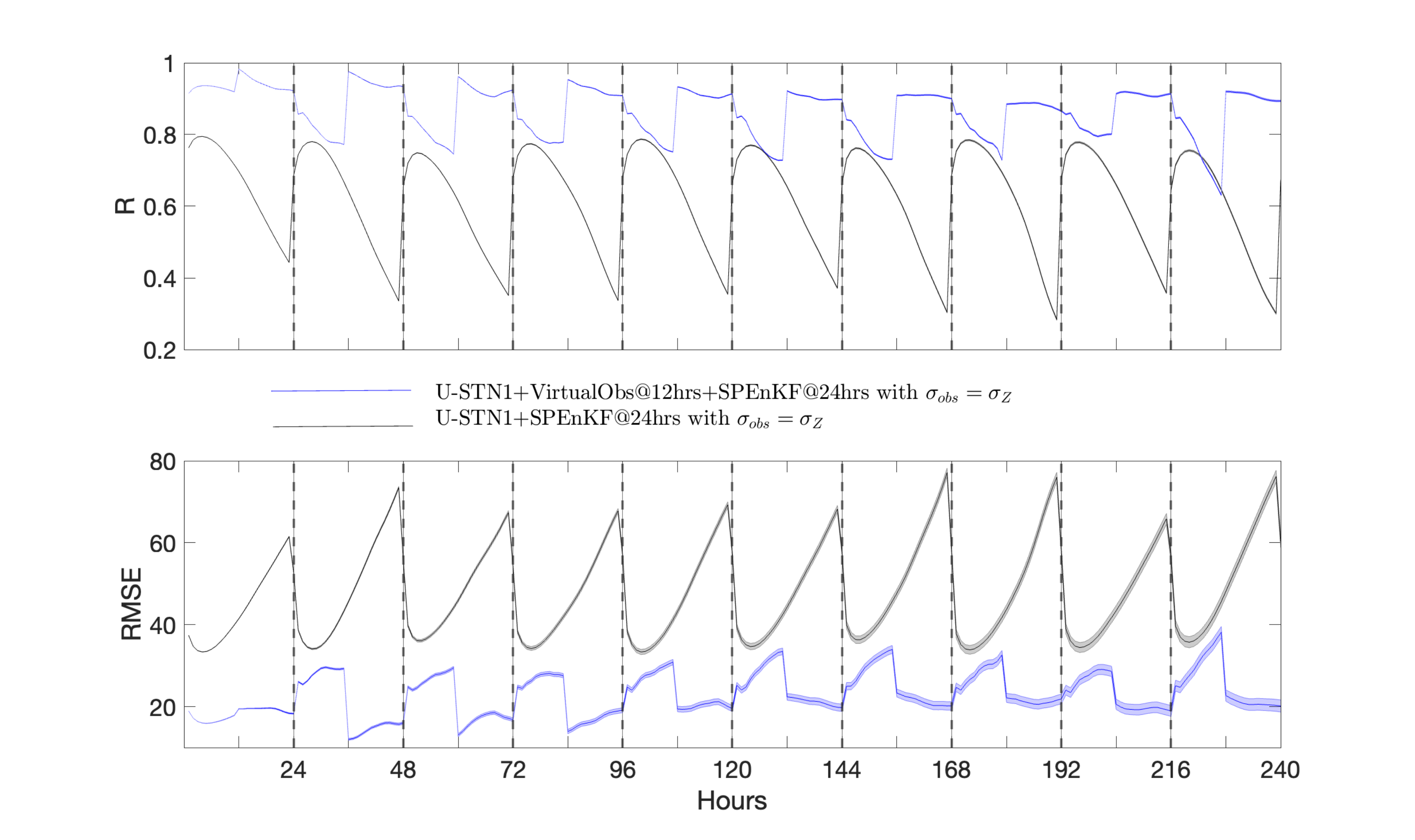}
\caption{\label{fig:DL_2DA_1} Same as Fig.~\ref{fig:DL_2DA_0.5} but with large observation noise, $\sigma_\text{obs}=\sigma_{Z}$.}
\end{figure}

Figure~\ref{fig:DL_2DA_0.5} compares the performance of the multi-time-step U-STNx+SPEnKF framework, which uses virtual observations from U-STN12, with that of U-STN1+SPEnKF, which was introduced in Section~\ref{sec:DD_with1DA}, for the case with $\sigma_\text{obs}=0.5\sigma_{Z}$. In terms of both RMSE and R, the multi-time-step U-STNx+SPEnKF framework outperforms the U-STN1+SPEnKF framework, as for example, the maximum RMSE of the former is often comparable to the minimum RMSE of the latter. Figure~\ref{fig:DL_2DA_1} shows the same analysis but for the case with larger observation noise $\sigma_\text{obs}=\sigma_{Z}$, which further demonstrates the benefits of the multi-time-step framework and use of virtual observations.  


The multi-time-step framework with assimilated virtual observations introduced here improves the forecasts of short-term intervals by exploiting the non-trivial dependence of the accuracy of autoregressive, data-driven models on time step size. While hourly forecasts of Z500 may not be necessarily of practical interest, the framework can be applied in general to any state variable, and can be particularly useful for multi-scale systems with a broad range of spatio-temporal scales. A similar idea was used in \citet{bach2021ensemble}, wherein data-driven forecasts of oscillatory modes with singular spectrum analysis and an analog method were used as virtual observations to improve the prediction of a Lorenz 96 chaotic dynamical system.

\conclusions[Discussion and Summary]
\label{sec:discussion}
In this paper, we propose three novel components for DDWP frameworks to improve their performance. These components are: 1) a deep spatial transformer in the latent space to preserve equivariances and encode the relative spatial relationships of features of the spatio-temporal data in the network architecture, 2) a stable and inexpensive ensemble-based DA algorithm to ingest noisy observations and correct the forecast trajectory, and 3) a multi-time-step algorithm, in which the accurate forecasts of a DDWP model that uses a larger time step are assimilated as virtual observations into the less accurate forecasts of a DDWP that uses a smaller time step, thus improving the accuracy of forecasts at short intervals. 

To show the benefits of each component, we use downsampled Z500 data from ERA5 reanalysis and examine the short-term forecast accuracy of the DDWP framework. To summarize the findings:
\begin{enumerate}
    \item As show in Section~\ref{sec: results_withoutDA} for noise-free initial conditions (no DA), the equivariance-preserving model, U-STN12, which uses a deep spatial transformer and $\Delta=12$~h, outperforms U-NET12, for example, extending the average prediction horizon (when ACC reaches 0.6) from 3.75~days (U-NET12) to 5.5~days (U-STN12). Examining examples of the spatio-temporal evolution of the forecasted Z500 patterns demonstrate that as expected, U-STN better captures phenomena such as wavebreaking, which involve spatial rotations and scalings. We further show in Section~\ref{sec:VirtualObs} based on other metrics that with the same $\Delta t$, U-STN outperforms U-NET. These results demonstrate the benefits of adding deep spatial transforms to convolutional networks such as U-NETs.
    \item As shown in Section~\ref{sec:DD_with1DA}, an SPEnKF DA algorithm is coupled with the U-STN1 model. In this framework, the U-STN1 serves as the forward model to data drivenly generate a large ensemble of forecasts in each DA cycle (24~h), when noisy observations are assimilated. Because U-STN1 is computationally inexpensive, for a state vector of size $D$, ensembles with $2D+1=4097$ members are easily generated in each DA cycle, leading to stable, accurate forecasts without the need for localization or inflation of covariance matrices involved in the SPEnKF algorithm. The results show that DA can be readily coupled with DDWP models when dealing with noisy initial conditions. The results further show that such coupling is substantially facilitated by the fact that large ensembles can be easily generated with data-driven forward models.
    \item As shown in Section~\ref{sec:VirtualObs}, the autoregressive DDWP models (U-STN or U-NET) are more accurate with larger $\Delta t$, which is attributed to the nonlinear error accumulation over time. Exploiting this trend and the ease of coupling DA with DDWP, we show that assimilating the forecasts of U-STN12 into U-STN1+SPEnKF as virtual observations in the middle of the 24~h DA cycles can substantially improve the performance of U-STN1+SPEnKF. These results demonstrate the benefits of the multi-time-step algorithm with virtual observations.   
\end{enumerate}
Note that to provide proof-of-concepts, here we have chosen specific parameters, approaches, and setups. However, the framework for adding these 3 components is extremely flexible, and other configurations can be easily accommodated. For example, other DA frequencies, $\Delta t$, U-NET architectures, or ensemble-based DA algorithms could be used. Furthermore, here we assume that the available observations are noisy but not sparse. The gain from adding DA to DDWP would be most significant when the observations are noisy and sparse. Moreover, the ability to generate O(1000) ensembles inexpensively with a DDWP would be particularly beneficial for sparse observations for which the stability of DA is more difficult to achieve without localization and inflation~\citep{asch2016data}. The advantages of the multi-time-step DDWP+DA framework would be most significant when multiple state variables, of different temporal scales, are used, or more importantly, when the DDWP model consists of several coupled data-driven models for different sets of state variables and processes~\citep{reichstein2019deep,schultz2021can}. Moreover, while here we show that ensemble-based DA algorithms can be inexpensively and stably coupled with DDWP models, variational DA algorithms could be also used, given that computing the adjoint for the DDWP models can be easily done using automatic differentiation. 

The DDWP models are currently not as accurate as operational NWP models \citep{weyn2020improving,arcomano2020machine,rasp_2020_resnet,schultz2021can}. However, they can still be useful through generating large forecast ensembles \citep{weyn2021sub} and there is still much room for improving DDWP frameworks, for example using the three components introduced here as well as using transfer learning, which has been shown recently to work robustly and effectively across a range of problems~\citep[e.g.,][]{ham2019deep, chattopadhyay2020super,subel2020_sgs,guan2021LES}. 

Finally, we point out that while here we focus on weather forecasting, the three components can be readily adopted for other parts of the Earth system, such as ocean and land, for which there is a rapid growth of data and need for forecast/assimilation~\citep[e.g.,][]{kumar2008land,kumar2008integrated,yin2011ensemble,edwards2015regional,liang2019using}.

\codedataavailability{All codes used in this study are publicly available at \url{https://github.com/ashesh6810/DDWP-DA}. The data are available from the WeatherBench repository at \url{https://github.com/pangeo-data/WeatherBench}.} 



\noappendix       




\appendixfigures  

\appendixtables   


\authorcontribution{A.C., M.M., and K.K. designed the study. A.C. conducted research. A.C. and P.H. wrote the manuscript. All authors analyzed and discussed the results. All authors contributed to writing and editing of the manuscript.} 

\competinginterests{The authors declare that they have no conflict of interest.} 

\begin{acknowledgements}
We thank Jaideep Pathak, Rambod Mojgani, and Ebrahim Nabizadeh for helpful discussions. This work was started at National Energy Research Scientific Computing Center (NERSC) as a part of A.C.'s internship in the summer of 2020 under the mentorship of M.M. and K.K., and continued as a part of his PhD work at Rice University under the supervision of P.H. This research used resources of NERSC, a U.S. Department of Energy Office of Science User Facility operated under Contract No. DE-AC02-05CH11231. A.C. and P.H. were supported by ONR grant N00014-20-1-2722 and NASA grant 80NSSC17K0266. A.C. also thanks the Rice University Ken Kennedy Institute for a BP HPC Graduate Fellowship. E.B. was supported by the University of Maryland Flagship Fellowship and Ann G. Wylie Fellowship, and by Monsoon Mission II funding (Grant IITMMMIIUNIVMARYLANDUSA2018INT1) provided by the Ministry of Earth Science, Government of India.
\end{acknowledgements}







 \bibliographystyle{copernicus}
 \bibliography{example.bib}

\end{document}